# Flux pinning characteristics in cylindrical ingot niobium used in superconducting radio frequency cavity fabrication

**Asavari S Dhavale[1], Pashupati Dhakal[2], Anatolii A Polyanskii[3], and Gianluigi Ciovati[2]**
[1] Accelerator and Pulse Power Division, Bhabha Atomic Research Center, Mumbai 400085, India
[2] Thomas Jefferson National Accelerator Facility, Newport News, VA 23606, USA
[3] National High Magnetic Field Laboratory, Florida State University, Tallahassee, FL 32310, USA

E-mail: dhakal@jlab.org

**Abstract.** We present the results of from DC magnetization and penetration depth measurements of cylindrical bulk large-grain (LG) and fine-grain (FG) niobium samples used for the fabrication of superconducting radio frequency (SRF) cavities. The surface treatment consisted of electropolishing and low temperature baking as they are typically applied to SRF cavities. The magnetization data were fitted using a modified critical state model. The critical current density $J_c$ and pinning force $F_p$ are calculated from the magnetization data and their temperature dependence and field dependence are presented. The LG samples have lower critical current density and pinning force density compared to FG samples which implies a lower flux trapping efficiency. This effect may explain the lower values of residual resistance often observed in LG cavities than FG cavities.

**PACS** 74.25.Ha, 74.25.Sv, 74.25.Wx

## 1. Introduction

Over the past decade, particle accelerators have been relying on superconducting radio-frequency (SRF) cavity technology to achieve high accelerating gradients with reduced losses [1]. SRF cavities are mostly made of high-purity (residual resistivity ratio > 300), fine-grain (ASTM 5) bulk niobium. Since 2005, large-grain (grain size of few cm$^2$) Nb discs directly sliced from ingots became a viable option to fabricate SRF cavities with performance comparable to that achieved by standard fine-grain Nb cavities [2]. Large-grain Nb could be, on a large scale, cheaper than fine-grain material. The performance of SRF cavities made of bulk Nb is often limited by a sharp increase of the RF losses starting at peak surface magnetic fields, $B_p$, greater than about 90 mT. This phenomenon is often referred to as "high field Q-slope" or Q–drop [3]. A low-temperature (~ 120 °C for 12–48 h) baking (LTB) in either Argon or ultra-high vacuum (UHV) is often found to be beneficial in reducing the Q–drop. Preliminary theoretical and experimental studies indicate that magnetic vortices pinned near the surface are one among the possible causes for the Q–drop [4]. Magnetic vortices can be produced in SRF Nb cavities because of the imperfect shielding of the Earth's magnetic field or thermoelectric currents during cavity cool down across the critical temperature. Magnetic flux can be pinned at material defects, such as grain boundaries, dislocations or clusters of impurities [5, 6].

Flux pining characteristics of niobium alloys such as NbTi and Nb$_3$Sn have been studied extensively in the past because of their use in superconducting magnets, where artificial pinning centers are introduced to reduce the dissipation due to the movement of vortices, but fewer reports exist on the pinning properties of single-crystal or polycrystalline high purity niobium used in SRF applications.

In order to study the pinning properties of Nb, DC magnetization measurements were carried out in the temperature range of 2 K to 8 K on large-grain as well as fine-grain niobium samples treated by electropolishing (EP) and low-temperature (120 °C) baking (LTB). These treatments are commonly used in the surface preparation of SRF cavities. From measurements of the magnetization curves at different temperatures, the temperature dependence of the field of first flux penetration $H_{ffp}$, and of the upper critical field, $H_{c2}$ were obtained. The critical current density as a function of the magnetic flux density was calculated from the irreversible magnetization curve using either the Bean model [7, 8, 9] or a modified critical-state model as proposed by Matsushita and Yamafuji for superconductors with low Ginzburg-Landau parameter, $\kappa$ [10]. The field dependence of the pinning force was analyzed using the standard pinning model for type-II superconductors [11, 12, 13]. In addition, semi-quantitative measurements of surface pinning characteristics were done at 2 K on the same samples by measuring the change of AC penetration depth during a linear ramp of the DC magnetic field applied parallel to the surface, between 0 T and 1 T and then back to 0 T.

**2. Experimental Set-Up**

*2.1. Sample Preparation*

The large-grain (LG) samples, labeled A to D were machined from Nb ingots of different purity as shown in table 1, from CBMM Brazil, while the fine-grain (FG) sample was machined from a solid rod from Wah Chang, USA. The samples used for this study are 6 mm in diameter and 120 mm in length with a 2 mm diameter concentric channel drilled down to about 115 mm. The inner channel was intended as a cooling channel since the RF properties of the sample's outer surface could be measured by inserting it into a "pill-box" cavity [14, 15]. This will be part of future studies. Prior to the experiments discussed in this contribution the LG samples were treated as follows [16]:

I. About 180 μm material removal by Buffered Chemical Polishing (BCP) with HF:HNO$_3$:H$_3$PO$_4$ = 1:1:1
II. Heat treatment at 600 °C/10 h in a UHV furnace
III. ~ 24 μm material removal by BCP 1:1:2
IV. Baking in UHV at 100 °C/12 h, 120 °C/12 h, 140 °C/12 h and 160 °C/12 h. About 10 μm were etched by BCP 1:1:2 after each bake.

The FG sample was treated as follows, prior to the experiments reported in this contribution:

I. About 65 μm material removal by Buffered Chemical Polishing (BCP) with HF:HNO$_3$:H$_3$PO$_4$ = 1:1:1
II. Heat treatment at 800 °C/2 h in a UHV furnace
III. ~ 140 μm material removal by BCP 1:1:1
IV. Heat treatment at 600 °C/10 h in a UHV furnace
V. Post-purification heat treatment at 1250 °C for 3 h using Ti as solid state getter
VI. ~ 100 μm material removal by BCP 1:1:1

After the post-purification heat treatment the grain size of the FG samples increased to about 1 mm

Here, we report the results from the measurements carried on the samples which were subjected the following surface and heat treatments:

I. ~ 50 μm material removal by electropolishing (EP) with HF:H$_2$SO$_4$ = 1:10 acid mixture.
II. Baking in UHV at 120 °C/48 h

**Table 1.** Contents ppm (per weight) of the main interstitial impurities from the different Nb ingots and RRR obtained from the samples' thermal conductivity at 4.2 K.

| Sample | Ta (ppm) | H (ppm) | C (ppm) | O (ppm) | N (ppm) | RRR |
|---|---|---|---|---|---|---|
| A | 1295 | 2 | <10 | 21 | 10 | 62 |
| B | 1310 | 2 | <10 | 9 | 3 | 164 |
| C | 603 | 4 | <10 | 14 | 9 | 159 |
| D | 644 | 3 | <10 | 7 | 7 | 118 |
| FG | <100 | <3 | <20 | <40 | <20 | 280 |

*2.2. Magnetization and Surface Pinning Measurement Setup*

A single-coil magnetometer [17] was built at Jefferson Laboratory to measure the DC magnetization of the hollow cylindrical Nb samples. Two heaters are clamped at the sample ends (distance ~ 70 mm) and two Cernox temperature sensors are also clamped in between (distance ~ 34 mm). A pick-up coil (~150 turns, length ~30 mm, and 0.28 mm diameter copper wire) is inserted in the center of the sample as shown in Fig. 1. The sample is clamped on a flanged copper block. The sample assembly is inserted in a copper tube that has a stainless steel Conflat flange brazed on one end and a copper flange on the other end. Indium wire provides the vacuum sealing between the two copper flanges. The assembly is bolted to a vertical test stand concentric to a custom built superconducting magnet from Cryomagnetics, Inc. that provides an axial magnetic field up to 1 T. The magnetic field homogeneity is 0.01 % over the length of the pick-up coil. The inside of the Cu tube is evacuated to a pressure lower than $10^{-4}$ mbar.

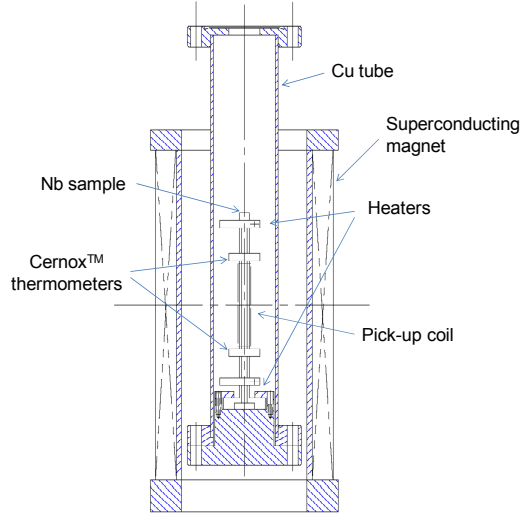

**Figure 1.** Schematic of the experimental set up.

The sample is zero field cooled (ZFC) from room temperature to 2 K. The DC magnetization measurement was carried out by linearly ramping up the magnetic field of the magnet from 0 to $H_{dc}$ at a ramp rate of 3.87 mT/s. The induced voltage in the pick-up coil is recorded with a nanovoltmeter (Model 2182, Keithley, Inc). The single-coil measurement depends critically on the regularity of the field sweeping. For this purpose a computer controlled magnet power supply programmer (Model 410, American Magnetics, Inc.) was used. The magnetization, $M$, as a function of the applied field, $H_{dc}$ is calculated by using the following equation [10]:

$$M(H_e) = \left(\frac{-1}{1-N_d}\right) \int_0^{H_{dc}} \frac{V(H') - V_n}{V_s - V_n} dH' \qquad (1)$$

where $V_s$ and $V_n$ indicate the voltages in the superconducting and the normal conducting state respectively and $N_d$ is the demagnetization factor (~0.007).

The measurements are carried out at various temperatures from 2–8 K. The temperature of the sample is changed by using the top and bottom heaters. The temperature over the sample length was monitored using calibrated Cernox temperature sensors. The power to the heaters was controlled by the proportional-integral-differential control loop of a temperature controller (Model 332, Lakeshore, Inc.) to maintain the temperature of the sample within a range of ± 0.1 K about the set point, $T_{set}$. After each magnetization measurement, the sample was warmed above the transition temperature and slowly cooled down to a new temperature $T_{set}$ in zero field.

By connecting the leads of the pick-up coil to a 30 nF capacitor outside the cryostat as part of a LC oscillator circuit, it is possible to measure the change in resonant frequency, $\Delta f$, as a function of the applied dc magnetic field cycled between 0–1T and back to 0T. The change in resonant frequency is proportional, through a geometric constant, to the change in penetration depth [18]. These measurements have been repeated at different temperatures, between 2 K and 8 K, and the samples are thermally cycled through $T_c$, in zero applied fields, after each measurement. The frequency of the oscillator at $H_{dc} = 0$ is ~270 kHz, which corresponds to a sampling depth of about 30 μm from the surface. The amplitude of the AC magnetic field produced by the coil is about 17 mG. The same oscillator circuit is used to measure the critical temperature of the sample by measuring the frequency change while slowly warming up the samples from 4.3 K to 10 K, in zero applied field.

## 3. Experimental Results
### 3.1 DC Magnetization and Penetration Depth

Figure 2 shows the results of the magnetization measurements carried out on samples D and FG at different temperatures after EP. The irreversibility in magnetization was observed consistently with the flux–line pinning of type–II superconductors. However, the irreversibility was also observed in the absence bulk pinning [19]. No significant dependence of the $M(H_{dc})$ curves on the magnetic field ramp rate, between 1.29 mT/s and 3.87 mT/s, was found. Although not shown here, similar results are found for the other samples in this field and temperature ranges. As shown in figure 2(b), the magnetization curves of the FG sample show an unusual curvature in the mixed state which might be related to the nature of flux pinning. In case of LG sample, flux jump was observed at the magnetic field ~ 0.3 T at 2 K which is believed to be due to the sudden redistribution of vortices caused by the thermo-magnetic instabilities [20]. However, the flux jump is not present at higher temperatures.

The $H_{ffp}$ ($H_{c1}$, for reversible magnetization) is extracted as the value of the applied field at which the magnetization curve deviates from the perfect diamagnetism, i.e., the position at which the magnetization curves deviate from the straight line as the external magnetic field is ramped up from zero to $H_e$ [21]. The temperature dependence of the $H_{ffp}$ and upper critical field $H_{c2}$ (extracted from the criteria $M(H_{c2}) \sim 0$, or the crossover from diamagnetic to paramagnetic state) is plotted as shown in insets of figure 2 for both samples D and FG, subjected to EP surface treatment. We fitted these data using the empirical relations

$$H_{ffp}(T) = H_{ffp}(0)\left[1 - \left(\frac{T}{T}\right)^2\right] \quad (2)$$

$$H_{c2}(T) = H_{c2}(0)\left[1 - \left(\frac{T}{T}\right)^2\right] \quad (3).$$

These expressions seem to reproduce the observed $H_{ffp}$ and $H_{c2}$ temperature dependence reasonably well. We have estimated the zero temperature $H_{ffp}$ and $H_{c2}$ to be 0.188 ± 0.005 T and 0.440 ± 0.010 T for sample-D, and 0.187 ± 0.003 T and 0.420 ± 0.008 T for sample-FG. Table 2 summarizes the critical fields values at 0 K obtained by fitting experimental data by Eqs. (2) and (3) for all samples. The estimated value of $H_{ffp}$ is consistent with values of $H_{c1}$ reported in the literature for high-purity niobium [22, 23]. Also, the estimated values of $H_{c2}$ are comparable to the recently reported values in niobium after BCP surface treatments [24]. These experimental results show the little or no influence LTB on the bulk superconducting properties.

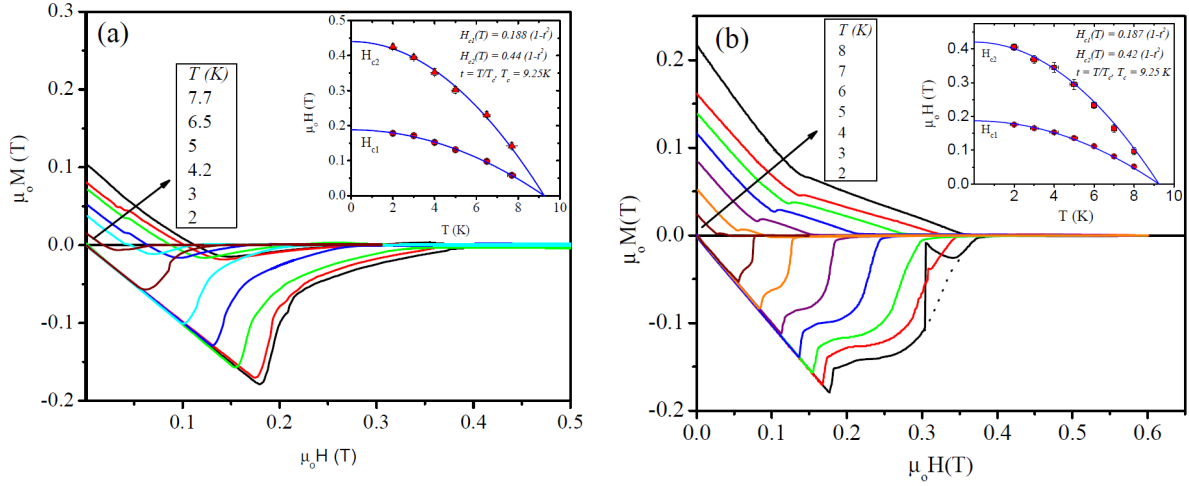

**Figure 2.** Magnetization curves of samples D (a) and FG (b) at different temperatures, measured after EP. The insets show the temperature dependence of critical fields. The solid lines are the calculated using (2) and (3).

**Table 2.** Critical fields values at 0 K obtained by fitting the data with (2) and (3).

| Sample | EP | | | EP+LTB (120 °C for 48 h) | | |
|---|---|---|---|---|---|---|
| | $\mu_0 H_{ffp}(0)$ (mT) | $\mu_0 H_{c2}(0)$ (mT) | $T_c$ (K) | $\mu_0 H_{ffp}(0)$ (mT) | $\mu_0 H_{c2}(0)$ (mT) | $T_c$ (K) |
| A | 187±4 | 418±8 | 9.25±0.01 | 190±7 | 427±8 | 9.22±0.02 |
| B | 183±3 | 411±10 | 9.12±0.02 | 185±4 | 421±8 | 9.26±0.02 |
| C | 194±5 | 443±11 | 9.21±0.02 | 193±11 | 445±12 | 9.24±0.01 |
| D | 188±5 | 440±10 | 9.21±0.03 | 192±9 | 439±11 | 9.24±0.02 |
| FG | 187±3 | 420±8 | 9.34±0.04 | 191±10 | 420±7 | 9.27±0.03 |

Figure 3 shows the change in resonant frequency of the LC oscillator as a function of external applied magnetic field at different temperatures for sample–C and FG, which is proportional to the change in penetration depth. This will allow us to determine the surface $H_{ffp}$, $H_{c2}$ and $H_{c3}$. When the external magnetic field is ramped up, no flux is penetrating and therefore there is no change in the oscillator's resonant frequency. Once the external field reached $H_{ffp}$, the flux starts to penetrate and the penetration depth increases gradually. The dependence of penetration depth above $H_{ffp}$ depends on the surface barrier (shielding current) as well as the surface pinning. At $H_{c2}$, the bulk sample becomes stable in its normal state, whereas the surface superconductivity still exists up to $H_{c3}$ [25]. With decreasing field, the penetration depth is reversible down to $H_{c2}$ and the irreversibility observed below $H_{c2}$ depends on the surface treatment. The larger the irreversibility near $H_{ffp}$ is, the stronger the surface barrier. The irreversibility is also expected due to fluxoids which are not in equilibrium with the external magnetic field because of surface pinning caused, for example, by NbO precipitates [26]. Another possible cause of the irreversibility between $H_{ffp}$ and $H_{c2}$ is the slight inhomogeneity of the order parameter, caused by the combination of the induced and diamagnetic screening currents, which is different in increasing and decreasing fields [27]. Table 3 summarizes the surface critical field values within ~30 μm depth from the surface for all samples after EP and EP followed by LTB heat treatments at 2 K. As shown, the surface critical field $H_{c3}$ (or the ratio $H_{c3}/H_{c2}$) increases due to the LTB. This is can be explained by the reduction of the electron mean free path due to diffusion of impurities during LTB [19, 25].

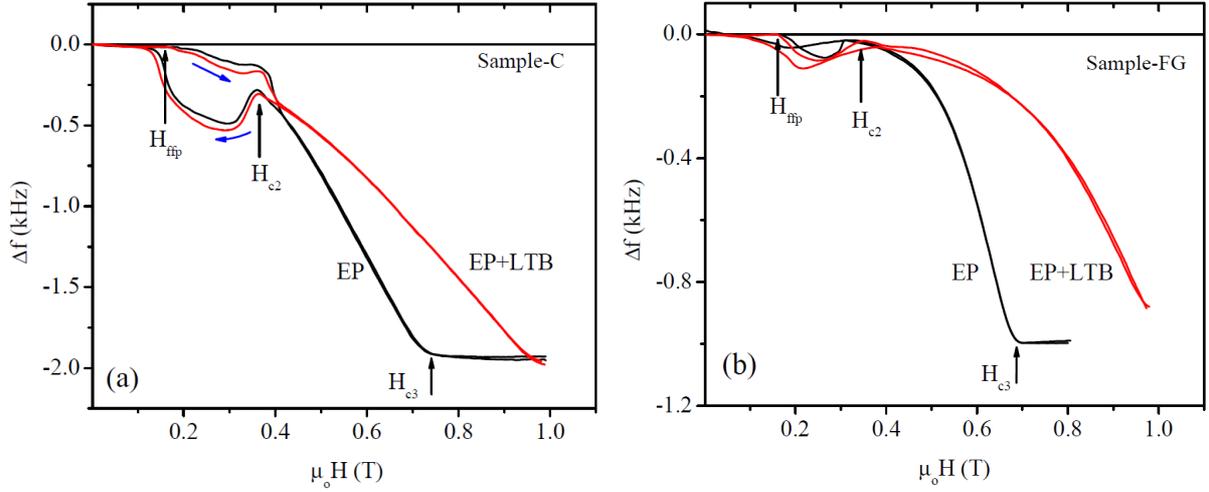

**Figure 3.** Frequency change as a function of the applied DC magnetic field measured at 2 K after different surface and heat treatments for sample−C (a) and sample−FG (b). The arrows on the plots indicate the position of the surface fields $H_{ffp}$, $H_{c2}$, and $H_{c3}$.

**Table 3.** Critical field values measured within a ~ 30 μm depth from the surface for all samples after EP and EP followed by LTB (120 °C for 48 h) at 2 K.

| Samples | EP | | | EP+LTB (120 °C for 48 h) | | |
|---|---|---|---|---|---|---|
| | $\mu_0H_{ffp}$ (mT) | $\mu_0H_{c2}$ (mT) | $\mu_0H_{c3}$ (mT) | $\mu_0H_{ffp}$ (mT) | $\mu_0H_{c2}$ (mT) | $\mu_0H_{c3}$ (mT) |
| A | 178±10 | 384±10 | 736±15 | 173±7 | 373±11 | 766±12 |
| B | 170±6 | 336±7 | 705±12 | 173±7 | 365±10 | >1000 |
| C | 160±8 | 333±6 | 753±10 | 165±6 | 353±9 | ~1000 |
| D | 165±7 | 345±6 | 710±13 | 166±6 | 347±9 | 745±12 |
| FG | 168±10 | 358±9 | 689±15 | 171±5 | 362±8 | >1000 |

*3.2 Magneto-optical Measurements on FG Samples*

In order to further understand the unusual curvature in the magnetization measurement of the fine-grain sample in the mixed state, six 2.5 mm thick samples were cut from the center portion of the rod by wire electro-discharge machining to observe the pattern of flux penetration using the magneto-optical (MO) technique [28]. After cutting, the samples were etched for 2 min by BCP 1:1:2, removing about 6 μm of material. The samples have a "doughnut" shape with an estimated demagnetization factor of about 0.51. The pattern of flux penetration was measured by MO for two out of the six samples. One sample, labeled #3, showed fairly uniform flux penetration and remnant field as the external magnetic field was cycled between 0 and 120 mT and back to 0 T, at a temperature of 5.4 K. The other sample, labeled #5, showed non-uniform flux penetration and remnant field distribution as the applied magnetic filed was cycled above $H_{ffp}$ and back to zero at different temperatures, between 6 K and 8 K. Figure 4 shows the optical image of the top surface of sample #5 and the corresponding MO image at 6 K, 124 mT, indicating a non uniform flux penetration, with some preferential flux entry through some of the grain boundaries. Evidence for preferential flux penetration at grain boundaries has been reported in a detailed study of high-purity Nb samples by MO imaging [29] and the results showed, in particular, that the occurrence of this phenomenon strongly depends on the orientation between the applied field and the grain boundary plane. The $H_{ffp}$ values measured by MO at different temperature in the range 6.5 – 9 K are consistent with those obtained for the full rod with the single-coil magnetometer. The MO results suggest that the unusual curvature in the mixed state in the $M(H)$ plot of figure 2 (b) could be due to the inhomogeneous flux distribution in the volume sampled by the coil.

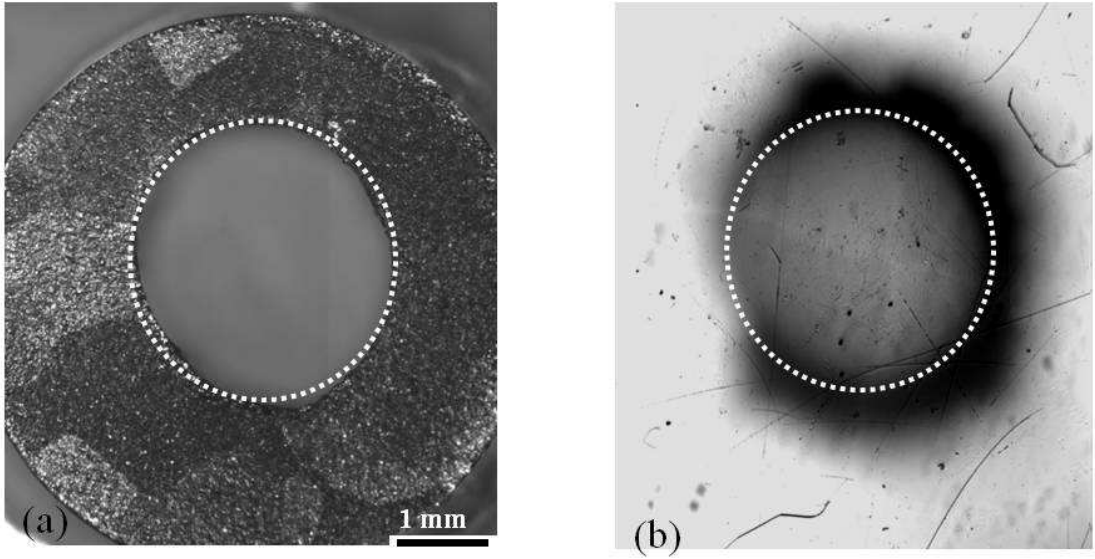

**Figure 4.** Optical image (a) of the cross-section of sample #5 cut from the FG hollow rod sample and MO image (b) after zero-field cooling of the same sample at 6 K and 124 mT. The dark areas in (b) indicate flux-free regions of the sample.

## 4. Calculation of Critical Current Density and Pinning Force

*4.1 Critical State Models*

Bean's model is one of the most widely used critical state models to describe the field and current distribution in a superconductor [7, 8, 9]. The critical state describes a current distribution throughout the superconductor determined by the effectiveness of material inhomogeneities in pinning a flux distribution against Lorentz forces [30]. This allowed us to determine the field dependent critical current density $J_c(B)$ in terms of the measured values of magnetization, under the assumption of full flux penetration as

$$J_c(B) = \frac{3}{2}\Delta M(B)\left(\frac{R_{out}^2 - R_{in}^2}{R_{out}^3 - R_{in}^3}\right) \qquad (4)$$

where $\Delta M(B)$ is the hysteresis when the externally applied field is increased or decreased about $B/\mu_0$. Here, $R_{out}$ is the outer radius and $R_{in}$ is the inner radius of the sample. In the critical state of the superconductor the current density is equal to the critical value $J_c$ and the pinning force exactly balances the Lorentz force. Therefore the pinning force density $F_p$ can be calculated as:

$$F_p(B) = J_c(B)B \qquad (5)$$

Even though the Bean model successfully explained the critical state of high $\kappa$ type-II superconductor, it deviates for the low $\kappa$ and weakly pinned superconductors where the diamagnetic contribution to the critical state is significant. Much work has been done to understand the origin of the experimentally observed irreversible magnetization in low $\kappa$ type-II superconductors [31, 32, 33, 34, 35]. In all those models, several assumptions and empirical parameters were used to obtain a good fit of the experimental data, but there is no clear consensus on which one provides the most accurate physical description of the magnetic properties of low $\kappa$ type-II superconductors. For this study, we considered the modified critical state model proposed by Matsushita and Yamafuji. The model is described in details in Refs. [10, 13]. The magnetic field and flux density profiles in a cylindrical sample, $R_{in} \leq r \leq R_{out}$ sample are calculated numerically and the magnetization is obtained as

$$M = -\mu_0 H_e \pm \frac{B_{c2}^2}{\alpha(R_{out}-R_{in})} \int_{H(r=R_{in})}^{H_e} \left[\frac{B(H)}{B_{c2}}\right]^{2-\gamma} \left[1-\frac{B(H)}{B_{c2}}\right]^{-\delta} dH \qquad (6)$$

where the positive sign is for increasing external field and the negative sign is for decreasing field. $\alpha$, $\gamma$ and $\delta$ are fit parameters resulting from the widely used empirical law was assumed for the pinning force density [36]:

$$F_p(B) = \alpha \left(\frac{B}{B_{c2}}\right)^\gamma \left(1-\frac{B}{B_{c2}}\right)^\delta \qquad (7)$$

For decreasing external field, $H_e$, the following dependence $H(r)$ is proposed in the model:

$$H(R) = \frac{a}{b}\left(1-e^{-bR}\right) \qquad \text{for } 0 \leq H_e \leq H_{c1} \qquad (8)$$

with $a$ and $b$ being two additional fit parameters, $R = r + r_0$, where $r_0$ is chosen as $H = 0$ at $R = 0$. The parameter $a$ corresponds to the critical current density at $H_e = 0$.

The experimental data for the magnetization as a function of reduced magnetic field ($H/H_{c2}$) for sample-B after EP and LTB treatment are shown in figure 5. The symbols in figure 5 are obtained from a calculation of the magnetization using Eq.(6) with the parameters listed in table 4, showing an excellent agreement with experimental data. The data for all LG samples could be fitted well with Eq. (6). The modified critical state model cannot provide a good fit of the magnetization curve for the FG sample in increasing field, between $H_{ffp}$ and $H_{c2}$, due to the unusual curvature of $M(H)$ in the mixed state which we attribute to a highly non-uniform flux distribution in the sample's volume, because of preferential flux entry at some of the grain boundaries.

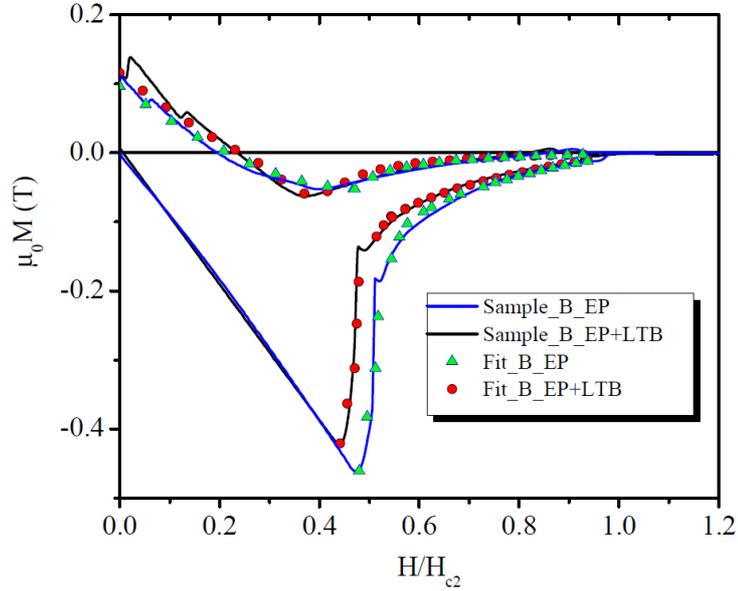

**Figure 5.** Magnetization as a function of reduced applied field ($H/H_{c2}$) for sample-B after EP and LTB treatments at 2 K. The solid lines are the experimental data where as the points are the results of numerical calculations using the modified critical state model with the parameters given in table 4.

**Table 4.** The parameters used to fit the experimental magnetization curves using the modified critical state model of Matsushita and Yamafuji [10]

| Fitting Parameters | Sample-B | | Sample-FG | |
|---|---|---|---|---|
| | EP | EP+LTB | EP | EP+LTB |
| $\alpha$ (N m$^{-3}$) | $8.38\times10^6$ | $8.08\times10^6$ | $2.38\times10^7$ | $2.13\times10^7$ |
| $\gamma$ | 1 | 1 | 0.5 | 0.5 |
| $\delta$ | 1.25 | 1.25 | 1.25 | 1.25 |
| $\beta$ | 0.08 | 0.08 | 0.05 | 0.05 |
| $a$ (A m$^{-2}$) | $1.29\times10^8$ | $1.37\times10^8$ | $3.5\times10^8$ | $3.37\times10^8$ |
| $b$ (m$^{-1}$) | $4.26\times10^8$ | $4.31\times10^2$ | $5.13\times10^2$ | $5.06\times10^2$ |

*4.2 Critical Current Density and Pinning Force*

Figure 6 shows a plot of the critical current density at 2 K as a function of reduced magnetic flux density ($b=B/B_{c2}$) for sample-D after to EP, calculated using Bean's model (Eq. (4)) from the experimental data. Using the modified critical state model described above, it is also possible to calculate the critical current density from Eqs. (6) and (10), as shown in figure 5. The critical current density calculated using Bean's critical state model and Matsushita and Yamafuji's modified critical state model are in good agreement at high magnetic flux density, however a large deviation is visible at low field ($b \sim 0.13$). This is consistent with previous studies showing that Bean's model underestimates $J_c$ at low magnetic flux densities even for high $\kappa$ superconductors, compared to other critical state models which better describe magnetization data [37, 38].

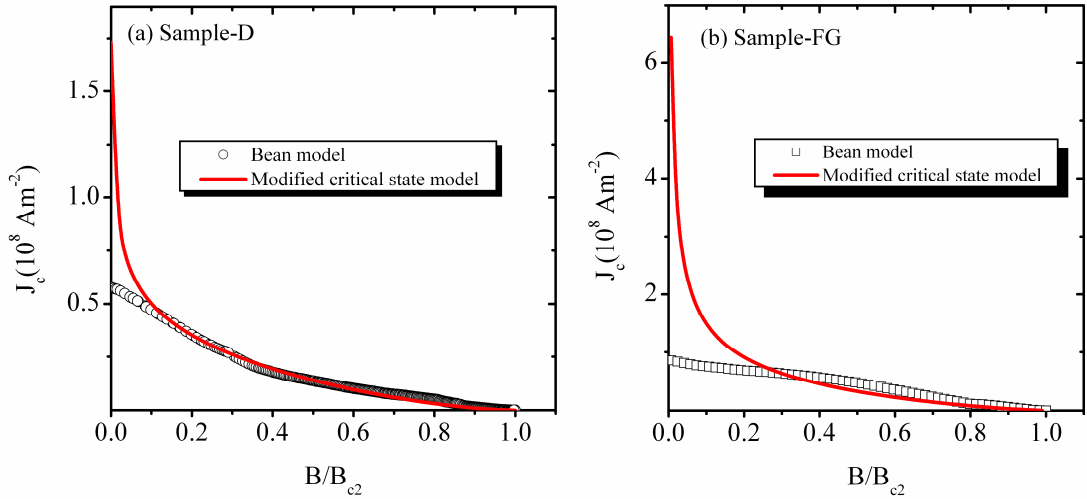

**Figure 6.** Critical current density as a function of reduced magnetic flux density ($b=B/B_{c2}$) at 2 K for (a) sample-D (b) sample-FG after EP surface treatment. The critical current density is calculated from magnetization data using Bean's model and the modified critical state model described in the text.

*4.3 Flux Pinning Mechanism*

The pinning force depends on temperature, applied magnetic field and microstructure of the sample under consideration. Any dislocation networks, grain boundaries or surface precipitation with sizes of the order of the coherence length (about 40 nm in Nb) can act as a pinning center. These pinning centers are randomly distributed within the sample and are capable of pinning one or multiple vortices at a particular site. In view of this scenario it is not possible to give an exact mathematical expression for the basic pinning force involved but one can study the scaling of the pinning force with various parameters.

In a simple case where vortices are randomly distributed and the applied magnetic field is not very large (near $H_{ffp}$), each pinning site can be considered as independent and the total pinning force is just the summation due to all the pinning sites. Considering $N_p$ as the number of pinning sites in a unit volume and $f_p$ as the elementary pinning force involved, the total pinning force density has a simple dependence

$$F_p \propto N_p f_p \tag{9}$$

As the magnetic field is increased, the spacing between the vortices is reduced and therefore an attractive or repulsive interaction between vortices results in a pinning force that is less than the mere summation of force due to individual vortices. In short, the average superconducting order parameter and hence the average magnetic induction is changed with variation in the applied magnetic field. Thus, the experimental curves show a maximum at particular value of reduced magnetic flux density, $b$. An empirical scaling law of the pinning force per unit volume, $F_p$ is given by [30]

$$F_p \propto H_{c2}^m f(b) \approx H_{c2}^m b^\gamma (1-b)^\delta \tag{10}$$

In order to understand the intrinsic pinning mechanism of the system, the normalized pinning force density $F_p/F_p(max)$, calculated from (5), was plotted against the reduced magnetic flux density $b$ as shown in figure 7 for samples D and FG after EP. The critical current density from Bean's model was used in Eq. (5). As it is shown in figure 7, $F_p/F_p(max)$ as a function of $b$ follows the same curve at all temperatures, suggesting that the basic pinning force in a sample is unique and depends on the microstructure (grain size). The normalized pinning force has the peak value at a reduced magnetic flux density of $0.34 \pm 0.02$ for sample-D whereas it is $0.46 \pm 0.02$ for the FG sample. For many materials containing second phase, or grain boundaries, the maximum in normalized pinning force occurs about 0.33, for the dislocation pinning on the other hand it can vary between 0.25–0.8 depending on the deformation structures [32]. In this case, the large shift of the peak in normalized pinning force to the higher value is observed in the FG sample.

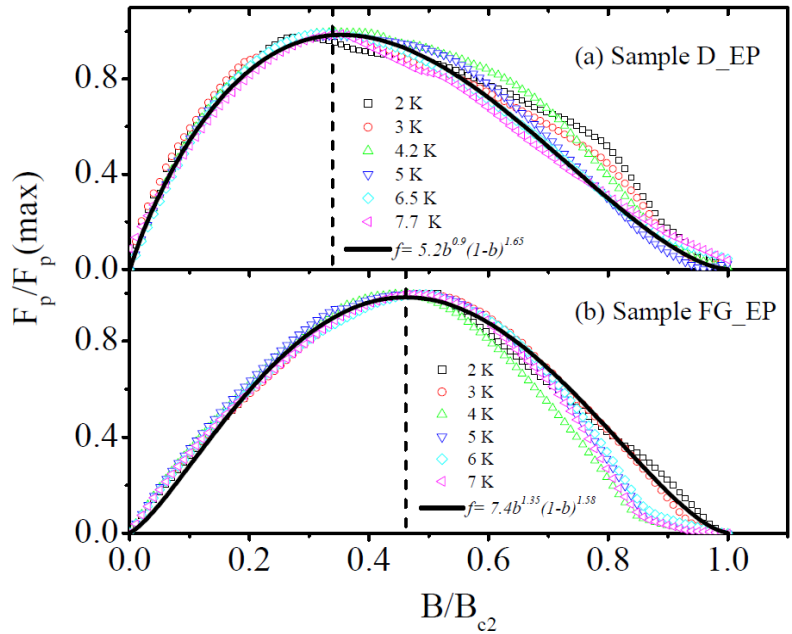

**Figure 7.** Normalized pinning force versus the reduced magnetic flux density $b=B/B_{c2}$ at different temperatures for samples D (a) and FG (b) after EP surface treatments. The solid line is an approximate fit to the experimental data.

We have also plotted the approximate fit to the normalized pinning force with the function $F_p/F_p(\max) = Ab^\gamma(1-b)^\delta$ with the parameters $A = 5.2$, $\gamma = 0.9$ and $\delta = 1.65$ for sample-D, whereas $A = 7.4$, $\gamma = 1.35$ and $\delta = 1.58$ for sample-FG. The value of peak reduced magnetic flux density can be calculated to be $b_{peak} = \gamma/(\gamma+\delta)$, which yields $b_{peak} = 0.36$ for sample-D and $b_{peak} = 0.46$ for sample-FG. This scaling expression fits well the experimental data up to $b_{peak}$, however some deviations are visible closer to $B_{c2}$.

The temperature dependence of $F_p$ is defined by the first term in Eq. (10) while terms containing reduced magnetic flux density define the basic pinning force. A plot of the maximum pinning force, $F_p(max)$ at different temperatures and the corresponding upper critical field, $H_{c2}(T)$ as shown in figure 8. The slope of the straight line fit gives the index $m$ and hence the dependence of $F_p$ on $H_{c2}(T)$. In all the LG samples we studied, $m \sim 1.5 \pm 0.01$ and in case of FG, $m = 1.82 \pm 0.02$.

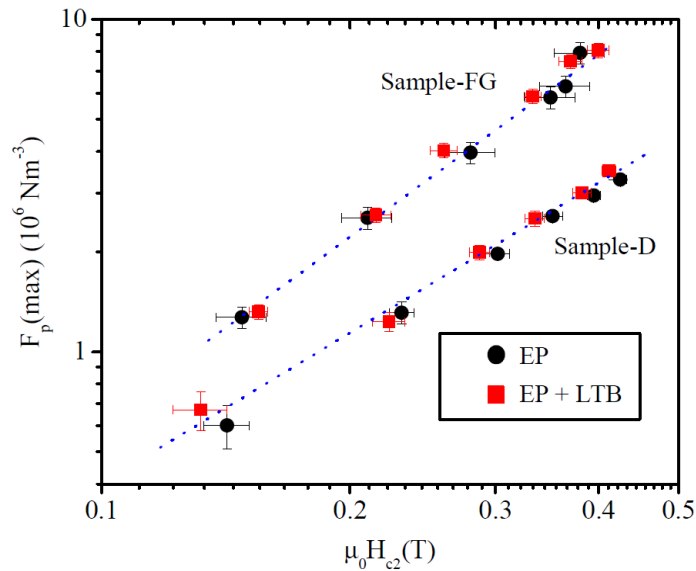

**Figure 8.** Scaling of the maximum pinning force with the upper critical field for sample-D after EP and LTB. The dotted line is the result of the fit $F_p(max) \sim H_{c2}^m$, yielding $m = 1.50 \pm 0.01$ for sample-D and $m = 1.82 \pm 0.02$ for sample-FG.

## 5. Discussion

The results on the critical current density showed no correlation with the RRR of the samples and therefore with the bulk impurity concentrations. Measurements of the depth profile of the critical current density on low purity (RRR ~ 40) Nb samples indicated $J_c$ values at 4.2 K up to about $5\times10^{10}$ A m$^{-2}$ at the surface, leveling off to less than about $2\times10^7$ A m$^{-2}$ at a depth of about 10 μm [39]. Similar results were obtained on single-crystal samples of the same purity. Oxygen depth profiles calculated from values of $H_{c2}$ measured at different depths indicated oxygen concentrations above 400 at. ppm within a 30 μm depth from the surface. From those results, it was inferred that material inhomogeneities, such as oxygen and hydrogen precipitates, dominated the pinning mechanism. For the higher purity Nb which is now available for SRF cavity fabrication, the concentration of impurities have been significantly reduced and the influence of grain boundary on pinning properties is more evident, as shown by the large-grain and fine-grain samples measured for this study. Impurities depth profiling by secondary ion mass spectrometry showed O, N and C levels below 180 at. ppm [40]. On the other hand, hydrogen concentrations of several tens of atomic percent have been measured near the surface [41] as hydrogen is readily absorbed into Nb during chemical treatments.

Penetration depth measurements as a function of an applied external magnetic field showed a broadening of the transition to $H_{c3}$ after LTB. Such broadening can occur due to surface roughness or material inhomogeneities. Since baking at 120 °C does not affect the surface roughness, the measurement results suggest an increase of surface inhomogeneities. Oxygen diffusion within a 50 nm depth from the surface had been proposed as a possible explanation of the baking effect [42] but experiments trying to test the validity of the model showed contradicting results (see for example the review in Ref. [38]).

In relation to the application to SRF cavity fabrication, the results presented in this contribution indicate that large-grain Nb would be less efficient in pinning magnetic flux during the cavity cool-down, compared to fine-grain Nb, because of the lower $J_c$. This would result in reduced RF losses (higher $Q_0$-value) for large-grain cavities. Results on direct measurements of trapped flux in fine-grain and large-grain Nb sample were recently reported [43] and showed a lower trapping efficiency in large-grain Nb, consistently with our conclusion. Furthermore, comparison of several cavities RF test results at DESY, Germany [44], and Jefferson Lab [45] indicate lower residual resistance and higher $Q_0$-values in large-grain, compared to fine-grain cavities which underwent similar surface treatments.

## 6. Conclusions

DC magnetization measurements carried out on large-grain Nb samples of different purity and one fine-grain Nb sample of high-purity at various temperatures allowed obtaining the field and temperature dependence of the critical current density and pinning force on such samples. Using the modified critical state model by Matsushita and Yamafuji, the irreversible magnetization was calculated showing good agreement with the experimental data. The calculated $J_c$ and $F_p$ of large grain samples (A-D) are lower than the FG, as expected because of the fewer grain boundaries. Even though the LG samples have different RRR values, the magnetic properties of these large grain samples do not depend on the bulk impurity concentrations. Because of the lower $J_c$-value, it is expected that large-grain niobium would have lower pinning efficiency than fine-grain Nb. This should result in lower RF losses and therefore higher $Q_0$-values for large-grain cavities. Recent results on cavities and samples confirmed these expectations. The improvement of $Q_0$ at 2 K and accelerating gradients of about 20 MV/m is important in order to reduce the cryogenic losses in superconducting continuous wave accelerators for a variety of applications.

**Acknowledgements**

The authors would like to acknowledge Dr. G. R. Myneni, Dr. K. C. Mittal, and A. V. Gurevich for several interesting discussions. The authors would also like to acknowledge P. Kushnick and M. Morrone of JLab for technical support.

This manuscript has been authored by Jefferson Science Associates, LLC under U.S. DOE Contract No. DE-AC05-06OR23177. The U.S. Government retains a non-exclusive, paid-up, irrevocable, world-wide license to publish or reproduce this manuscript for U.S. Government purposes.